\documentclass[superscriptaddress,twocolumn]{revtex4-1}

\usepackage{amsmath}
\usepackage{graphicx}
\usepackage{dcolumn}
\usepackage{latexsym}
\usepackage{color}
\usepackage{subfigure}
\usepackage{comment}
\usepackage{nicefrac}

\usepackage[ulem=normalem]{changes}
\usepackage{float}

\usepackage{xcolor}
\usepackage{soul}

\begin{document}

\title{Hybrid Double Quantum Dots: Normal, Superconducting, and Topological Regimes}

\author{D.~Sherman}
\affiliation{Center for Quantum Devices and Station Q Copenhagen,
Niels Bohr Institute, University of Copenhagen, Copenhagen, Denmark}
\author{J.~S.~Yodh}
\affiliation{Center for Quantum Devices and Station Q Copenhagen,
Niels Bohr Institute, University of Copenhagen, Copenhagen, Denmark}
\author{S.~M.~Albrecht}
\affiliation{Center for Quantum Devices and Station Q Copenhagen,
Niels Bohr Institute, University of Copenhagen, Copenhagen, Denmark}
\author{J.~Nyg\aa rd}
\affiliation{Center for Quantum Devices and Station Q Copenhagen,
Niels Bohr Institute, University of Copenhagen, Copenhagen, Denmark}
\author{P.~Krogstrup}
\affiliation{Center for Quantum Devices and Station Q Copenhagen,
Niels Bohr Institute, University of Copenhagen, Copenhagen, Denmark}
\author{C.~M.~Marcus}
\affiliation{Center for Quantum Devices and Station Q Copenhagen,
Niels Bohr Institute, University of Copenhagen, Copenhagen, Denmark}

\maketitle

\textbf{Epitaxial semiconductor-superconductor hybrid materials are an excellent basis for studying mesoscopic and topological superconductivity, as the semiconductor inherits a hard superconducting gap while retaining tunable carrier density \cite{Chang_hardgap}. Here, we investigate double-quantum-dot devices made from InAs nanowires with a patterned epitaxial Al two-facet shell \cite{Peter} that proximitizes two gate-defined segments along the nanowire. We follow the evolution of mesoscopic superconductivity and charging energy in this system as a function of magnetic field and voltage-tuned barriers. Inter-dot coupling is varied from strong to weak using side gates, and the ground state is varied between normal, superconducting, and topological regimes by applying a magnetic field. We identify the topological transition by tracking the spacing between successive cotunneling peaks as a function of axial magnetic field \cite{Sven} and show that the individual dots host weakly hybridized Majorana modes.}

Proximitized semiconductors with an induced hard superconducting gap \cite{Chang_hardgap} offer new possibilities to explore the interplay between charging energy and mesoscopic superconductivity. Interest in hybrid systems has intensified recently due to proposals \cite{Oreg,Lutchyn} to use hybrid materials to create one-dimensional topological systems that support Majorana end modes, including concrete schemes for fusing and braiding Majorana modes using single and branched proximitized nanowires \cite{Aasen}. Quantum state control via braiding makes use of predicted non-Abelian exchange statistics of Majorana modes, defining a path towards topological quantum information processing \cite{Freedman,Nayak}.  Majorana modes have been identified in single-constriction \cite{LeoScience,MingtangNano,Das,Churchill} and single-island devices \cite{Sven, Yazdani}. However, more complex device geometries are required to advance the field toward future applications. 
 
\begin{figure}
\begin{center}
\hspace{0cm}
\includegraphics[scale=0.43]{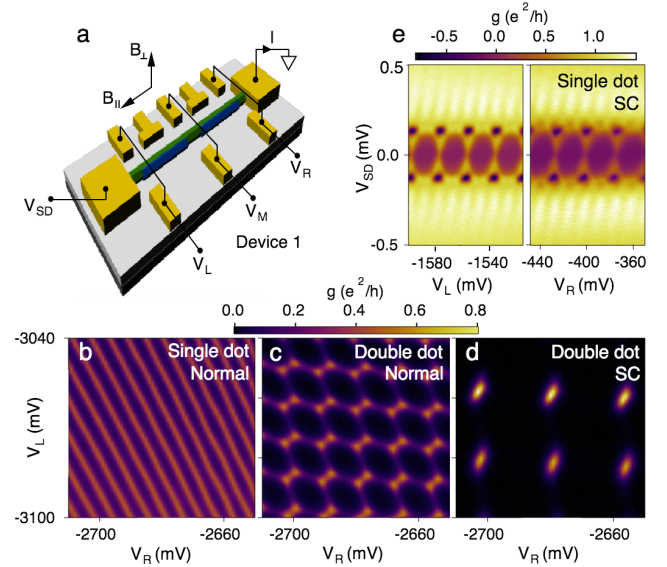}
\vspace{-4.7cm}
\caption{\label{fig:HiggsCond}(Color online) \textbf{Double quantum dot with controllable coupling and superconductivity}. \textbf{a}, 	Device schematic shows quantum dots defined by etching Al (blue) from the InAs (green) nanowires, with normal metal (Ti/Au) contacts and electrostatic side gates. Conductance, $g$, versus gate voltages $V_L$ and $V_R$, at zero source-drain voltage, $V_{SD}$ = 0 forms a two-dimensional charge stability diagram (CSD). \textbf{b}, Applying $B_\perp = 0.7$ T quenches superconductivity. At middle gate voltage $V_M = -5$ V, a single quantum dot is formed, with diagonal stripes corresponding to 1$e$-periodic Coulomb blockade peaks. \textbf{c}, Applying $V_M = -5.5$ V separates the dots, resulting in a honeycomb-pattern CSD. \textbf{d}, Lowering $B_\perp$ drives the system into the superconducting state, yielding  2$e$-periodic honeycomb vertices. \textbf{e}, Differential conductance, $g$, as a function of $V_{SD}$ and $V_L$ ($V_R$) at $V_M=-5$ V shows 2$e$-periodic Coulomb diamonds for $eV_{SD}<\Delta$ and 1$e$-periodic Coulomb diamonds for $eV_{SD}>\Delta$. Note that $V_L$ and $V_R$ tune the joint superconducting dot with similar efficiency.}
\vspace{-0.5cm}
\end{center}
\end{figure}

The minimal system to test Majorana fusion rules is a single nanowire with two topological islands coupled by a controllable central barrier \cite{Aasen}. In the scheme, a Majorana pair is initialized with the central barrier open. Then, when the barrier is closed, Majorana pairs in the two dots are projected into a superposition that can be read out with charge detectors by closing the two outer barriers. These and related experiments rely on a high degree of device tunability.  With this in mind, we explore the basic properties of hybrid superconducting double quantum dots, including a magnetic field driven transition from $2e$ to $1e$ periodicity of charge occupation of both dots. We then tune the system to the topological regime and detect Majorana end modes with weak splitting in a double-dot geometry using a characteristic inversion of Coulomb blockade peak spacing \cite{Sven}.

\begin{figure}
\begin{center}
\vspace{0cm}
\hspace{-0.45cm}
\includegraphics[scale=0.62]{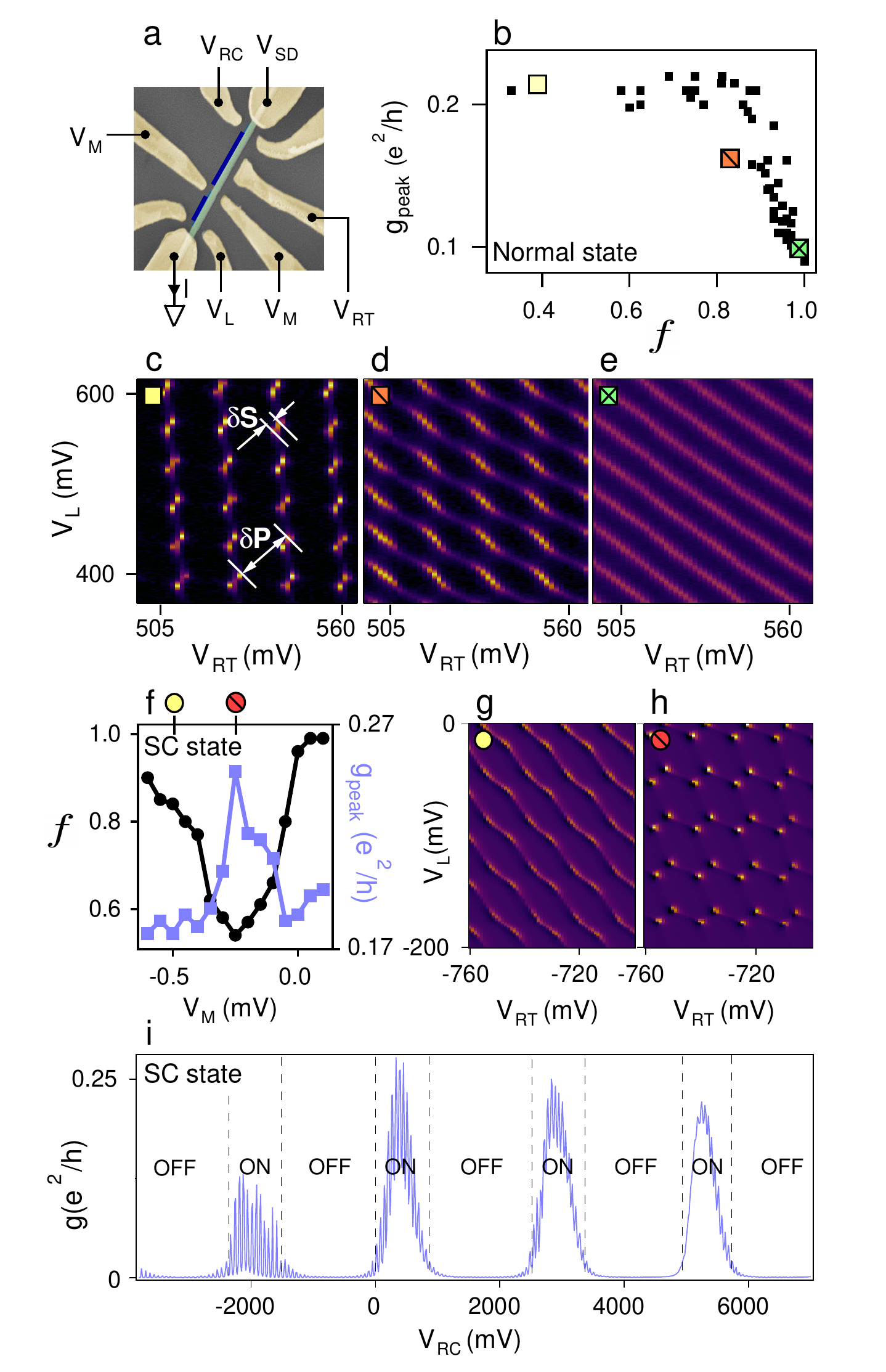}
\vspace{-0.5cm}
\hspace{0cm}
\caption{\textbf{Tuning inter-dot coupling. a,} Electron micrograph with false color of device 2 (similar to device 3) with epitaxial Al regions indicated schematically (blue). Here the patterned dots are 0.5$\mu$m (left) and 1.5$\mu$m (right). \textbf{b,} Peak conductance, $g_{\rm peak}$, as a function of fractional peak splitting, $f=2\delta S/\delta P$, obtained from device 2 in the normal state, with $B_\perp = 0.7$ T. Note that $g_{\rm peak}$ decreases as $f$ increases. \textbf{c-e,} CSDs with various $f$ values, controlled by $V_{M}$. $\delta S$ and $\delta P$ are shown in (c). \textbf{f,} $g_{\rm peak}$,  (blue, right side) and $f$ (black, left side) versus the middle side gate $V_M$, obtained from device 3 in the superconducting regime. Note the nonmonotonic behavior of $f$ versus $V_M$. \textbf{g-h,} CSDs for different $f$ values. \textbf{i,} Conductance $g$ versus $V_{RC}$, shows alternating on/off pattern, corresponding to the width and spacing of the right junction dot energy levels. The high-frequency oscillations are Coulomb peaks associated with the right patterned dot. Panels f-i were taken at finite DC bias, $V_{SD}=10~\mu$V.}
\label{fig:tun_thz}
\vspace{-0.6cm}
\end{center}
\end{figure}

InAs nanowires with wurtzite crystal structure and hexagonal cross section were grown by molecular beam epitaxy in the [0001] direction with a 10 nm  epitaxial Al shell on two facets \cite{Peter}. The Al was etched in a $\sim$100 nm section in the center of the wire, and from the ends of the wire using Transene D etch. The Al remaining on the two segments of nanowire defined the double dot. Normal metal (Ti/Au) electrostatic gates, patterned by e-beam lithography, were used to tune the tunnel barriers across the constrictions and the chemical potential of each of the designed quantum dots. Ohmic contacts at both ends of the wire were made by removing the native oxide from the nanowire with {\em in-situ} Ar milling then depositing normal metal (Ti/Au). Three devices were measured. Device 1 had equal segments of 1 $\mu$m each (Fig.~1a), and devices 2 and 3 had one short (0.5 $\mu$m) and one long (1.5 $\mu$m) segment (Fig.~2a).

Tuning the devices between the superconducting and the normal state is achieved by applying a magnetic field.  Tuning the double-dot coupling between fully coupled and fully decoupled is achieved by applying a voltage on the middle side gate, $V_M$. Conductance, $g$, at zero source-drain voltage, $V_{SD}$, as a function of left and right side gate voltages, $V_L$ and $V_R$, which constitutes the well known double-dot charge stability diagram (CSD) \cite{van_der_Weil}, was measured across these transitions. Applying a perpendicular (see Fig.~1a) magnetic field of $B_{\perp}=0.7$ T quenches the superconducting energy gap, $\Delta$, and allows resonant tunneling of single electrons across the device. With a transparent middle constriction (open middle valve), a single quantum dot is formed between the outer constrictions. The diagonal lines in Fig.~1b separate states with charge difference $e$ in the single large dot. Closing the middle valve breaks the system into two separated quantum dots, with a familiar honeycomb CSD (Fig.~1c). At vertices of the honeycomb, left and right dots are on resonance with one another and with the leads \cite{van_der_Weil}. 

Superconductivity is induced by lowering the magnetic field. Now the smallest charge unit that can traverse the device is a Cooper pair with a charge of $2e$, and consequently the period of the honeycomb pattern is doubled \cite{Tuominen,Bibow}, as seen in Fig.~1d. Reopening the middle barrier converts the device to a joint superconducting quantum dot. Differential conductance as a function of $V_{SD}$ and either $V_L$ or $V_R$ reveals evenly spaced Coulomb peaks with a periodicity of $2e$ for $eV_{SD}<\Delta$ and $1e$ for $eV_{SD}>\Delta$, where $\Delta \simeq 0.2$ meV is the induced superconducting energy gap of the InAs (see Fig.~1e). Note that each $2e$-periodic Coulomb peak terminates at $eV_{SD} \sim \Delta$ with an abrupt transition to a region of negative differential conductance, in agreement with recent studies of single proximitized quantum dots \cite{Andrew_parity,Sven}.

\begin{figure}
\begin{center}
\vspace{-0.65cm}
\hspace{-2cm}
\includegraphics[scale=0.64]{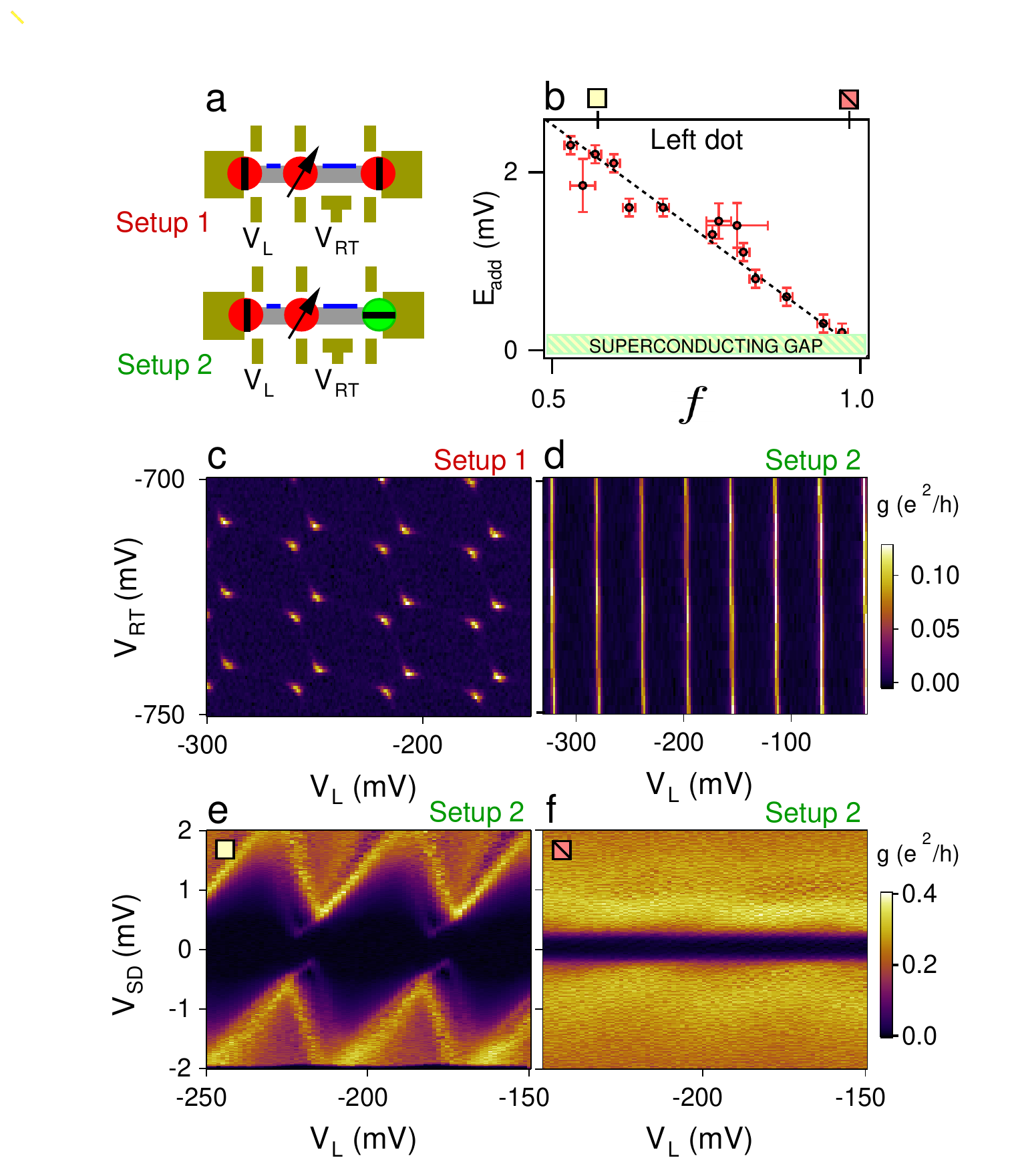}
\hspace{-1.7cm}
\vspace{0cm}
\caption{\textbf{Addition energy as a function of inter-dot coupling.} \textbf{a}, The two setups used to measure the addition energy, $E_{add}$, of the left dot as a function of its coupling to the right superconductor. Setup 1 has three closed valves. Here $f$ is measured in the normal state as a function of $V_M$. In setup 2 the right valve is open, quenching the charging energy of the right dot to well below $\Delta$. Here $E_{add}$ is measured as a function of $V_M$. \textbf{b,} $E_{add}$ as a function of $f$, as both quantities were mapped to $V_M$. \textbf{c,} A typical normal-state double-dot CSD from which $f$ is extracted \textbf{d,} CSD from setup 2. With the right right valve open, sweeping $V_{RT}$ has no effect while $V_L$ changes the number of electrons on the left dot. \textbf{e-f,} $V_{SD}$ versus $V_{L}$ for (e) relatively closed middle valve, $E_{add}>\Delta$, and (f) relatively open middle valve, $E_{add}<\Delta$. $E_{add}$ is defined as value of $eV_{SD}$ at the apex of the Coulomb diamonds.}

\vspace{-0.5cm}
\label{spectral}
\end{center}
\end{figure}

\begin{figure*}
\begin{center}
\hspace{-2cm}
\includegraphics[scale=0.75]{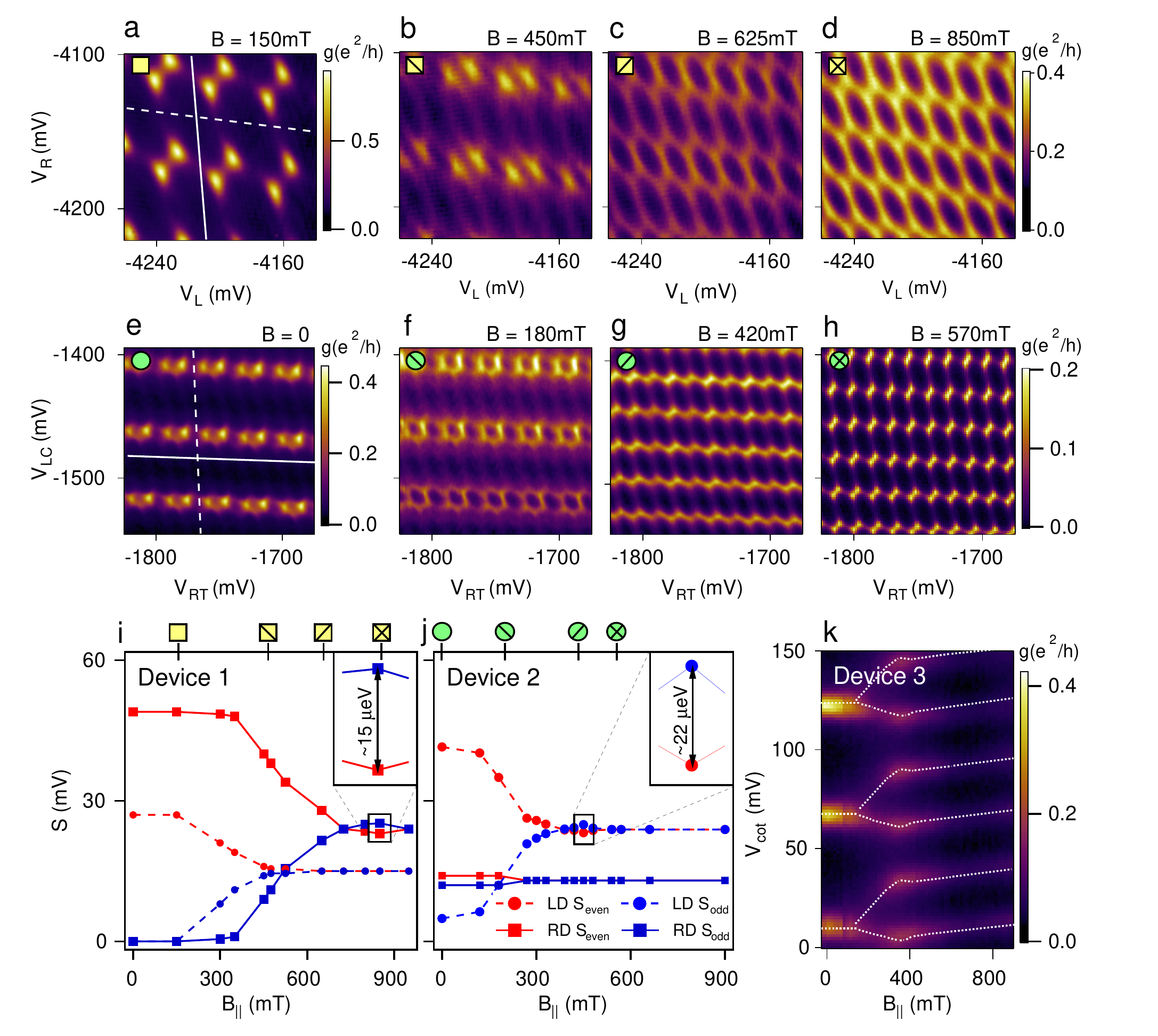}
\hspace{-2cm}
\vspace{-0.2cm}
\caption{\label{fig:HiggsCond}(Color online) \textbf{Signatures of Majorana modes via even-odd peak spacing along cotunneling lines.} \textbf{a-h}, Charge stability diagrams at increasing magnetic field. (a-d), Conductance, $g$ as a function of gate voltages, $V_R$ and $V_L$, for device 1 (see Fig.~1a) and (e-h) as a function of $V_{LC}$ and $V_{RT}$ for device 2 (see Fig.~2a). Cotunneling lines for the left (right) dot are shown in (a) and (e) as dashed (solid) white lines. \textbf{i}, Even-odd cotunneling peak spacing, $S$, of device 1 for the left (circles) and right (squares) quantum dots as a function of magnetic field. Data points were calculated by averaging along several cotunneling lines of the right and the left dots. Peaks from the right dot become evenly spaced around $B_{||} = 750$mT, above which they oscillate once about $S_{even} = S_{odd}$. Inset: zoom in of spacing oscillation with the extracted energy amplitude. \textbf{j}, same as (i) for device 2. In device 2, the oscillation is observed in the left dot. \textbf{k}, Differential conductance as a function of $B_{||}$ for device 3 along a cotunneling axis, $V_{\rm cot}$, corresponding to the left dot ($0.5$~$\mu$m).  Increasing $B_{||}$ halves the peak spacing from 2$e$ to 1$e$. The overshoot in peak spacing around $B_{||} = 350$mT is consistent with (i) and (j).}
\vspace{-0.5cm}
\end{center}
\end{figure*}

We next investigate tunablity of the middle constriction, which controls the inter-dot coupling, using devices 2 and 3. A useful metric of inter-dot coupling strength is the fractional peak splitting \cite{Waugh}, $f = 2\delta S/\delta P$, where $\delta S$ is the diagonal splitting measured between vertices and $\delta P$ is the distance between vertex pairs in a CSD. Both capacitive coupling and inter-dot tunneling rates determine $f$; however, opening the middle valve should increase the tunnel coupling exponentially faster than the capacitive coupling \cite{LPK1}. Neglecting capacitive coupling, when $f=1$, the inter-dot tunnel coupling is maximized, and the device behaves like a single quantum dot. Contrary to previous studies \cite{YONGJIE,WaughPRB,LPK1,Livermore,Waugh}, the differential conductance on the Coulomb peak, $g_{\rm peak}$, increases as the inter-dot coupling is reduced, as seen for instance in Figs.~2b-e. These results were obtained  in the normal state, by applying $B_\perp = 0.7$ T. Similar behaviour is obtained in the superconducting regime ($B = 0$), and pushing $V_M$ to either positive or negative values results in a nonmonotonic inter-dot coupling (Figs.~2f-h). Here, the observed spacing between conductance peaks is dominated by charging energies, as they exceed $\Delta$. We interpret the nonmonotonic coupling as a signature of an unintentional dot formed in the junction. The barrier dot modulates the coupling between the larger, patterned dots, periodic in gate voltage.  Note that alternating resonances were also obtained for the side constriction, as shown in Fig.~2i. Junction dots and resonances are frequently encountered in the etched regions of epitaxial hybrid structures \cite{Chang_hardgap, Larsen}. We speculate that the dependence of $g_{\rm peak}$ on $f$ is related to a change in the mean free path, namely the effective mean free path is larger in the isolated double dots than in the joint single dot. If the charging energy of the two patterned-dots, $E_C$, is smaller than $\Delta$ (as in Fig.~1), the transport dynamics through the middle valve are determined by three parameters: the width of the junction dot energy levels, $\Gamma$, the junction dot charging energy, $\tilde{E}_{C}$, and the induced superconducting gap enclosing the junction, $\Delta$  \cite{Silvano}. Evidently, in Fig.~1 the device is tuned to the strong coupling limit, $\Gamma > \Delta, \tilde{E}_C$, and when in the superconducting regime (\textit{e.g.,} Fig.~1d), only Cooper pairs can propagate through the device \cite{LafargeN}. Consequently, we can consider these constrictions as valves that can be continuously opened or closed by scrolling up or down the junction dot energy levels with a side gate ($V_M$). However, as one gradually decreases $V_M$ the effective sizes of the patterned left and right dots shrink. Thus, at a sufficiently negative voltage the effective areas of each of the dots are pushed far enough from the constriction to quench the overall conductance. 

Considering the constrictions as tunable valves allows us to explore the evolution of a superconducting quantum dot addition energy, $E_{add}$, as we gradually vary its coupling to a superconductor with little or no charging energy. This problem has been investigated theoretically in Ref.~\cite{Kamenev}. $E_{add}$, which in the present case is the sum of $\Delta$ and $E_C$ \cite{van_der_Weil}, is the energy needed to add a single charge unit to the dot. It is given by $eV_{SD}$ at the apex of the measured Coulomb diamonds. Here we employed two setups (Fig.~3a) to obtain the dependence of $E_{add}$ of the left dot on $f$, which is shown in Fig.~3b. Setup 1, in which all three valves are closed to form a double quantum dot, was used to obtain $f$ for a given $V_M$, with a typical CSD shown in Fig.~3c. In setup 2, we open the right valve, which effectively quenches the charging energy of the right dot. Consequently, sweeping $V_{RT}$ no longer affects the charge occupancy of the device (Fig.~3d). In this setup, $g$ versus $V_{SD}$ and $V_L$ was measured and $E_{add}$ of the left dot was extracted at various $V_M$ values. By mapping both $f$ and $E_{add}$ to $V_M$ we extract the parametric dependence of $E_{add}$ on $f$. The observed linear dependence of $E_{add}$ on $f$ has not been addressed theoretically to our knowledge. 
Note that closing the middle valve progressively increases $E_{add}$ to more than an order of magnitude above $\Delta$ (Fig.~3e). On the other hand, opening the middle valve reduces $E_{add}$ to $\Delta$ (Fig.~3f), indicating that $E_C$ is quenched, up to the accuracy of the measurement.

In the $2e$ state, the charging energy, $E_C$, of each dot is smaller than the energy of the lowest quasi-particle state, $E_0$, of that dot. Consequently, Cooper pairs are the only allowed charge carriers, and each successive honeycomb cell in the CSD corresponds to a state with a charge difference of $\pm 2e$. Increasing the magnetic field lowers $E_0$ by the Zeeman energy, causing the evenly spaced $2e$ periodic Coulomb peaks to  split into large-small spacing corresponding to even-odd dot charge occupancies \cite{Averin, Lafarge}. Consequently, the unit cells in the CSDs alternate in size based on the ratio $E_0/E_C$. Following even-odd peak spacing as a function of magnetic field was recently used to detect Majorana end modes in a single island geometry \cite{Sven,vanHeck}.

In a single Majorana island geometry, Majorana zero modes are expected to hybridize with an energy splitting that depends exponentially on the length of the island and oscillates with magnetic field \cite{Sarma, Kitaev, Rainis, Stanescu}. Indeed, such oscillations were recently observed by following the distance between successive Coulomb blockade peaks as a function of magnetic field in the topological regime with an energy amplitude, $A \propto e^{-L/\xi}$, where $L$ is the island length and $\xi$ is the superconducting coherence length in the topological regime \cite{Sven}. In a double Majorana island geometry, similar oscillations are expected, provided that either the two Majorana modes adjacent to the middle constriction are not hybridized when both dots are in the topological regime or that only one of the dots is tuned to the topological regime. Here we focus on the latter scenario, where only one dot is in the topological regime, so that we do not need to account for Majorana splitting due to finite coupling through the middle barrier. We rely on measuring the spacing between successive elastic cotunneling peaks within the double dot CSD. Peaks along an elastic cotunneling line occur when one of the dots is on resonance with the normal leads \cite{van_der_Weil}, so the initial and final energy states are equal. Measuring the distance between even, $S_{even}$, and odd, $S_{odd}$, cotunneling peaks of a single dot in a double dot system is equivalent to following Coulomb peak spacings in a single dot system. 

We observe the evolution of the elastic cotunneling peak spacing, $S$,  as a function of magnetic field applied parallel to the nanowire, $B_{\parallel}$. $V_{SD}$ versus $B_{\parallel}$ reveals a finite $\Delta$ that persists above $1$ T (not shown), in accordance with recent measurements on similar devices \cite{Sven}. Figure 4 shows several CSDs taken at different $B_{\parallel}$ values from devices 1 and 2. In device 1 (Figs.~4a-d), the zero field CSD is $2e$ periodic on both dots. In device 2 (Figs.~4e-h), the zero-field CSD exhibits an even-odd pattern for both dots. The difference between peaks, which is averaged over multiple cotunneling lines, gradually decreases with increasing $B_{\parallel}$, as seen in Fig.~4i(j) for device 1(2). The peak spacing is converted from gate voltage to energy units using the gate lever arm $\eta$, which was extracted independently from the Coulomb diamonds. The Majorana splitting amplitude $A= \eta (S_{even}-S_{odd})$, following Ref.~\cite{Sven}. At a higher $B_{\parallel}$, but still below the superconducting critical field, $A$ changes sign (Figs.~4i-j insets) with an amplitude of $|A_1| \simeq 15$~$\rm\mu$eV (right dot, length of $1\mu$m) and $|A_2| \simeq 22$~$\rm\mu$eV (left dot, length of $0.5$~$\rm\mu$m) for device 1 and device 2, respectively. These overshoot amplitudes, where the odd valley size exceeds the the even valley for the first time as the field is increased, are in rough agreement with values in Ref.~\cite{Sven} for the corresponding device lengths. We can also directly track the evolution of the Coulomb peaks along a cotunneling axis of one of the dots, $V_{\rm cot}$, as we continuously vary the magnetic field (Fig.~4k). Here, $V_{\rm cot}$ is a linear combination of the left and right gate voltages that follows a particular cotunneling line. In all three devices, the spacing of adjacent peaks reaches a constant $1e$ periodicity as $B_{\parallel}$ exceeds the superconducting critical field.

The ability to selectively tune the energy scales that set the ground state and the interaction of each of the dots is key to future application and fundamental studies of mesoscopic superconductivity, in particular topological superconducting devices. Future work on similar devices with the addition of charge sensors is a leading prototype system to examine Majorana modes manipulation, such as fusion rules. Braiding Majorana modes is expected to be feasible by adding a third branch, forming a T-shaped device with three patterned dots \cite{Aasen}.

\vspace{1cm}

\textbf{Acknowledgements} 

We gratefully acknowledge support from the Danish National Research Foundation and Microsoft Research. CMM acknowledges support from the Villum Foundation. 

\textbf{Author Contributions} 

P.K. and J.N. developed the nanowire materials. D.S. fabricated the devices. D.S., J.Y. and S.M.A. carried out the measurements with input from C.M.M. D.S. analyzed the data. D.S., J.Y. and C.M.M. wrote the paper. All authors discussed the results and commented on the manuscript. 

\textbf{Competing financial interests} 

The authors declare no competing financial interests.

\end{document}